\documentclass[journal,comsoc]{IEEEtran}
\usepackage[T1]{fontenc}
\usepackage{amssymb,amsfonts,amsthm,amsmath}
\usepackage{amsmath}
\usepackage{geometry}
\usepackage[cmintegrals]{newtxmath}
\usepackage{url}
\usepackage{color}
\usepackage {graphicx}
\usepackage{multirow}
\usepackage{cite}
\usepackage{mathtools}
\usepackage{bigstrut} 
\usepackage{subcaption}
\usepackage{hyperref}
\usepackage{algorithm}
\usepackage{caption}
\usepackage{algpseudocode}
\usepackage{tikz}
\usepackage{pgfplots}
\theoremstyle{definition}
\newtheorem{thm}{Theorem}
\newtheorem{lem}{Lemma}
\newtheorem{corrol}{Corollary}
\def\x{{\mathbf x}}

\def\w{{\mathbf w}}
\def\v{{\mathbf v}}

\def\e{{\mathbf e}}

\def\y{{\mathbf y}}

\def\n{{\mathbf n}}
\def\t{{\mathbf t}}
\def\x{{\mathbf x}}
\def\0{{\mathbf 0}}
\def\A{{\mathbf A}}
\def\I{{\mathbf I}}

\def\F{{\mathbf F}}

\def\H{{\mathbf H}}

\def\CN{{\mathcal{CN}}}

\def\E{{\mathbb E}}

\begin{document}
\title{An Optimized SLM for PAPR Reduction in Non-coherent OFDM-IM}
\author{\IEEEauthorblockN{{Sarath Gopi} and {Sheetal Kalyani}}\\ 
	\thanks{The paper is currently under review.}
}
\maketitle
\begin{abstract}
 In this letter, we propose a peak-to-average power ratio (PAPR) efficient non-coherent orthogonal frequency division multiplexing with index modulation (OFDM-IM). It is shown that the non-coherent OFDM-IM design, which minimizes PAPR, is a non-linear optimization problem. This can be visualized as the optimization of the phase factor in selected mapping (SLM) technique. Further, a special case is considered, where the inputs take only real values. We then show how to approximately solve it using simple linear integer  programming and explicitly quantify the gap between the approximate and the optimal solutions. A computationally efficient heuristic scheme is developed to obtain a suboptimal solution of the integer optimization problem.  Finally, our simulation results indicate the merits of the proposed schemes. 
\end{abstract}
\begin{IEEEkeywords}
	OFDM, index modulation, integer programming, PAPR, SLM.
\end{IEEEkeywords}

    \section{Introduction}
   Index modulation assisted OFDM (OFDM-IM) is a popular choice for next generation wireless communication systems due to its merits over conventional OFDM~\cite{8004416, 7469311}. It conveys the information through not only the conventional PSK/QAM signals, but the sub-carriers that bear these classical symbols also. Recently, a non-coherent version of OFDM-IM has been proposed, where the information is conveyed only through the position of sub-carriers that holds a non-zero amplitude~\cite{8091014, 8737925}. The advantage of non-coherent OFDM-IM is that the channel state information (CSI) is not required. 
    \par 
    A high value of peak-to-average power ratio (PAPR) is a critical issue in the case of OFDM systems causing non-linear distortion and hence resulting in degradation of the performance~\cite{6476061}.  Unfortunately OFDM-IM also suffers from high PAPR~\cite{7469311}. In general, PAPR reduction techniques for OFDM can be directly applied to OFDM-IM. For example, the partial transmit sequence method is extended to OFDM-IM in \cite{8417538}. There are a few attempts specific to OFDM-IM are also reported in literature.  In \cite{7891632}, the idle sub-carriers in OFDM-IM have been exploited to reduce PAPR. Two low complexity PAPR reduction techniques are detailed in \cite{8433654}. The former method is exploiting the idle sub-carriers in OFDM-IM, while the latter one uses the active constellation extension method and idle sub-carriers to achieve a low PAPR. The single carrier principle, which is a well known technique for reducing PAPR in OFDM, is extended to OFDM-IM in~\cite{choi2019single}.
    \par 
    Selected Mapping (SLM) is a well known PAPR reduction technique for OFDM \cite{543811}. A number of variants of SLM have also been reported in literature \cite{929598, 6425284}. In SLM, the same information can be conveyed by more than one signal and the one having the least PAPR is selected for transmission. SLM can be used with non-coherent OFDM-IM without side information, since it is insensitive to the phase. In this communication, we optimize SLM for PAPR reduction in non-coherent OFDM-IM. The optimized version is able to achieve significantly more reduction in PAPR than the conventional SLM. Our major contributions are: 
    \begin{enumerate}
    \item We propose an optimized SLM (OSLM) for PAPR reduction in non-coherent OFDM-IM. The idea is to replace the constant, say $\beta$, which represents the active sub-carrier in non-coherent OFDM-IM with a complex number $\beta e^{j\phi}$.  A non-linear optimization problem is developed to choose the phase factor $\phi$ so that PAPR of the resultant time series is minimized. 
    \item We also consider a special case of the proposed OSLM, in which $\phi = 0 / \pi$. In this case, an approximate solution of the optimization problem is obtained using linear integer programming. The gap between the approximate and original solution is quantified. Finally a low complexity iterative technique is proposed to obtain a  suboptimal solution of the proposed problem. 
    \item  Finally, we show that the maximum likelihood (ML) detector for the proposed PAPR reduced non-coherent OFDM-IM  is same as that of the conventional non-coherent OFDM-IM. Hence, the methods do not offer any BER performance degradation compared to the original non-coherent OFDM-IM.
   \end{enumerate}
   
    \section{System Model}
    \label{sysmod}
     In non-coherent OFDM-IM \cite{8091014}, the sub-carriers in an OFDM frame are partitioned into a number of clusters. In each of these clusters, a part of sub-carriers are loaded with a constant `$\beta > 0$' to distinguish it from the rest of sub-carriers, which carry a zero. The selection of these active sub-carriers, i.e., the sub-carriers that carry a non-zero value is based on the information bits to be transmitted. Or in other words, a non-coherent OFDM-IM conveys information through the selection of active sub-carriers.
     \par 
     Consider an OFDM frame having $N$ sub-carriers, which are split into $B$ clusters each having $L = \frac{N}{B}$ sub-carriers. In each of these $B$ clusters, $K$ out of $L$ sub-carriers are loaded with a non-zero value depending on the information bits, where there are $\binom{L}{K}$ choices for these active sub-carrier selection. Hence,  a non-coherent OFDM-IM frame can send $B\lfloor \log_2 \binom{L}{K} \rfloor$ bits.
    \par 
    Let $\x = \left\{x(k)\right\}_{k=0}^{N-1} = \left[\x_0^T, ..., \x_{B-1}^T \right]^T$ be an $N \times 1$ vector representing an OFDM frame.  Here, $\x_b = \left[x(bL), x(bL+1), ..., x((b+1)L-1) \right]^T$ is the $b^{th}$ cluster, where $b={0, 1, ..., B-1}$. This OFDM frame is converted into the corresponding time domain (TD) signal $\x^t$ by taking inverse Fourier transform. For defining PAPR, we consider the oversampled version of the TD signal, given below:
    \begin{align}
    x^t (n) = \frac{1}{\sqrt{N}} \sum_{k=0}^{N-1}x(k)e^{j\frac{2\pi n k}{NR}}, ~0 \le n \le NR-1  
    \label{p2eqn0}
    \end{align} 
    where $R$ is the oversampling factor. Equivalently, (\ref{p2eqn0}) can be written as $\x^t = \F_p^H\x$, where $\F_p$ is the $N \times NR$ partial Fourier matrix, which is constructed by taking first $N$ rows of an $NR \times NR$ discrete Fourier transform (DFT) matrix. Now, PAPR is defined as follows \cite{4446229}:
    \begin{small}
  \begin{align}
  \text{PAPR}\left(\x^t\right) = \frac{\underset{n}{\max}|x^t(n)|^2}{\E\left\{ |x^t(n)|^2\right\}}=\frac{\underset{0 \le n \le NR-1}{\max} |\{\F_p^H\x\}_n|^2}{\E\left\{\x^H\x\right\}}  
  \label{p2eqn1}
  \end{align}
   \end{small}
  where $\E\{.\}$ is the expectation operation. (\ref{p2eqn1}) is due to Parseval's Theorem \cite{papoulis2002probability}. The proposed PAPR reduction technique is discussed in subsequent sections, is developed based on definition (\ref{p2eqn1}). 
   \section{Optimized SLM (OSLM) for Non-coherent OFDM-IM}
   \label{prop_papr}
   The basic principle behind SLM \cite{543811} is to keep more than one signal for representing the same information of an OFDM fame. Conventionally these signals, which are having identical information, are generated by multiplying the OFDM-frame with different phase vectors. Finally, one signal with the lowest PAPR is selected for transmission and a side information indicating the corresponding phase vector is also communicated in order to recover the signal at the receiver. SLM can be used for non-coherent OFDM-IM without the side information. However, the performance of SLM is limited by the number of phase vectors used. Hence, we propose an optimized SLM, which is able to achieve the lowest possible PAPR for non-coherent OFDM-IM. In order to explain the principle of OSLM, Lemma ~\ref{lempaprnc} is stated and proved below.
   \begin{lem}
   	\label{lempaprnc}
  	For the conventional non-coherent OFDM-IM with $B$ clusters, each having $K$ non-zero sub-carriers, the PAPR is $BK$, irrespective of the information bits. The peak power occurs at the first time domain sample.
   	\begin{proof}
   		Let $\beta > 0$ be the non-zero entries of the conventional non-coherent OFDM-IM frame $\x$. Note that there are exactly $BK$ non-zero entries. Hence,	in (\ref{p2eqn1}), the denominator is always $\frac{BK\beta^2}{N}$. Also we have,
   		\begin{align}
   		|\F_p^H\x| = \frac{1}{\sqrt{N}}\sum_{k, x_k \ne 0}\left| \e^{j2\pi \frac{kn}{N}} \beta \right| \le  \frac{1}{\sqrt{N}} BK\beta.
   		\label{p2eqn01}
   		\end{align}
   		The equality in (\ref{p2eqn01}) occur for $n=0$. Therefore, ${\max}~|\F^H\x|^2 = \frac{1}{N}B^2K^2\beta^2$. This proves the lemma.
   	\end{proof}
   \end{lem}
   Therefore, from the proof of Lemma \ref{lempaprnc}, it is clear that the peak power of non-coherent OFDM-IM will occur at the first sample of the time series, which corresponds to the average value in the frequency domain. This average value can be reduced to zero, if there are equal number of $+\beta$s and $-\beta$s. However, this may increase the peak power in other samples. Hence, we propose to use a phase vector $\beta\{e^{j\phi_i}\}_i$ to represent the active sub-carriers. This is equivalent to multiply the conventional OFDM-IM frame with a phase vector. The phase vector is selected in such a way that PAPR of the resultant time series is minimized. It should be noted that unlike SLM, there is no fixed  set of phase vectors, rather it is computed on the fly and hence can achieve a much lower PAPR. 
   \par   
   The problem is formulated as the minimization of~(\ref{p2eqn1}) with respect to $\x$. However, we need to consider only the numerator of (\ref{p2eqn1}) to formulate the problem, since the denominator is fixed at $\frac{BK\beta}{N}$ irrespective of the fact that non-zero entries are $\beta e^{j\phi}$. Without loss of generality, let $\beta=1$.  The PAPR minimization problem can be stated as follows:
   \par 
      \begin{align}
       \underset{\x}{\min}  ~\underset{n=0, ..., NR-1}{\max}& |\{\F_p^H\x\}_n|^2 \nonumber \\
      \text{such that}~&x_i = e^{j\phi_i}, ~0 \le \phi_i \le 2\pi ~ \forall~i \in \Gamma,
      \label{p2eqn02}
      \end{align}
       where $\Gamma$ is the set of active sub-carriers, i.e., $\Gamma = \{ i | x_i \ne 0,\text{ for}~ i=0, 1, ..., N-1\}$. Note that $|\Gamma| = BK$. The optimization problem (\ref{p2eqn02}) can be written as:
   \begin{align}
   P1:~&  \underset{\Phi}{\min}  ~\underset{n=0, ..., NR-1}{\max}  ~\tilde{g}_n(\Phi) \nonumber \\
   \text{such that}~& 0 \le \phi_i \le 2\pi, ~ \forall \phi_i \in \Phi, i \in \Gamma 
   \label{p2eqn03}
   \end{align}
   where,
   \begin{small}
   \begin{align}
   \tilde{g}_n(\Phi) &= BK + \sum_{i \in \Gamma}\sum_{j \in \Gamma, j \ne i}\cos\left( \frac{2\pi n}{NR}(i-j) + \left(\phi_i-\phi_j\right)\right) \nonumber \\
   &= BK + g_n(\Phi)
   \label{p2eqn04}
   \end{align}
    \end{small}
    (\ref{p2eqn04}) is obtained by expanding $|\{\F_p^H\x\}_n|^2$. Note, $P1$ is a typical constrained non-linear minimax optimization problem. This can be solved directly or can be converted into a non-linear program and solved using standard techniques \cite{ charalambous1978efficient, zangwill1967algorithm}. In Section \ref{intSol}, a special case in which $\phi_i =0/ \pi, \forall ~i$ is considered. In this case $P1$ reduces to a linear integer programming, This is also computationally intensive approach. However, this formulation helps to devise a heuristic approach, which is computationally efficient and achieve a lower PAPR than the conventional SLM. 
    \section{Special Case with Real Solution}
    \label{intSol} 
    Consider the case of OSLM, in which $\phi_i$ can take only two possible values, i.e., $\phi =0/ \pi, \forall i$. This is equivalent to assigning $\pm 1$ to represents the active sub-carriers. In this case, the PAPR minimization problem can be expressed as:
    \begin{align}
    P2:~ \underset{\x}{\min}  ~\underset{j=0, ..., N-1}{\max}|& \{\F_p^H\x\}_j| \nonumber \\
    \text{such that}~&x_i \in \{+1,-1\}, ~ \forall~i \in \Gamma,
    \label{p2eqn2}
    \end{align}
or equivalently:
     \begin{align}
      &\underset{z, x_i, i \in \Gamma}{\min} ~z \nonumber \\
     ~\text{such that} ~&z \ge | \{\F_p^H\x\}_j|, ~\forall~j  \in \{0, 1, ..., N-1\}\nonumber \\
     &\x_i \in \{-1, +1\} ~\forall~i \in \Gamma.
     \label{p2eqn3}
     \end{align}
     (\ref{p2eqn3}) is an integer optimization problem with non-linear constraints involving complex norms. This can be solved as follows.
     \subsection{An Approximate Linear Integer Program}
 We have, for any complex vector $\v = v^R +j v^I$, the norm $|\v| = \sqrt{\left(v^R\right)^2+\left(v^I\right)^2}$  can be expressed as~\cite[(4)]{streit1986solution}:
       	 \begin{align}
    	 |v| = \underset{0 \le \theta \le 2\pi}{\max} v^R \cos \theta + v^I \sin \theta.
    	 \label{p2eqn4}
    	 \end{align} 
Now,  (\ref{p2eqn3}) can be rewritten by linearising the non-linear constraints using (\ref{p2eqn4}) as:
    \begin{align}
     P3:~&\underset{z, x_i, i \in \Gamma}{\min} ~z \nonumber \\
     ~\text{such that} ~& z \ge  \{\F_p^H\x\}_j^R \cos \theta+ \{\F_p^H\x\}_j^I \sin \theta, \nonumber \\
     &~\forall~j  \in \{0, 1, ..., N-1\}, 0 \le \theta \le 2\pi \nonumber \\
     &\x_i \in \{-1, +1\} ~\forall~i \in \Gamma,   
    \label{p2eqn5}
    \end{align}
    where $\{\F_p^H\x\}_j^R$ and $\{\F_p^H\x\}_j^I$ are the real and imaginary part of the $j^{th}$ component of the complex vector $\F_p^H\x$. Note, (\ref{p2eqn5}) is a linear integer optimization problem. However, in order to solve (\ref{p2eqn5}) using standard techniques, the parameter $\theta$ has to be discretized.   Now, we will state and prove Lemma~\ref{lemp2ubnorm}, which will give bounds on the norm defined in (\ref{p2eqn4}) using a discrete set of $\theta$s~\cite{streit1986solution}. This is then exploited to discretize the constraints in~(\ref{p2eqn5}) with respect to $\theta$. 
    \begin{lem}
    	\label{lemp2ubnorm}
    	Let  $\mathcal{D} = \left\{\theta_p = (p-1)\frac{2\pi}{P}, p= 1, ..., P\right\}$ be the set of discretized values for $0 \le \theta \le 2\pi$ and the discrete norm of the complex vector $v = v^R+j v^I$  in the set $\mathcal{D}$ is defined by
    	\begin{align}
    	|v|_{\mathcal{D} } = \underset{\theta \in \mathcal{D} }{\max}~ v^R \cos \theta + v^I \sin \theta
    	\label{p2cs3}
    	\end{align} 
    	Then an upper and lower bound on the norm defined in (\ref{p2eqn4}) can be written in terms of $|v_{\mathcal{D}}|$ as:
    	\begin{align}
    	|v|_{\mathcal{D} } \le |v| \le |v|_{\mathcal{D} }\sec \left(\frac{\pi}{P}\right)
    	\label{p2cs4}
    	\end{align}
    	\begin{proof}
    			The left inequality is clear from the fact that the discrete set ${\mathcal{D}}$ considers only $P$ discrete values of $0 \le \theta \le 2\pi$. The right inequality can be proved as follows.	
    			\par 
    			Let $\theta^*$ and $\theta^*_p$ be the value corresponding to the maximum in~(\ref{p2eqn4}) and~(\ref{p2cs3}), respectively. For the specified set $\mathcal{D}$, let $\theta^*_p = \theta^* + \delta$, where $-\frac{\pi}{P} \le \delta \le \frac{\pi}{P}$. Hence, using (\ref{p2cs3}), we can write, 
    			\begin{align}
    			|v|_{\mathcal{D} } & =  v^R \cos \left(\theta^* + \delta \right) + v^I \sin \left(\theta^* + \delta \right) \nonumber \\
    			&= |v|\cos \delta  - \left\{v^R \sin \theta^* - v^I \cos \theta^* \right\} \sin \delta
    			\label{p2cs5}
    			\end{align}
    			Note that, the maximum in~(\ref{p2eqn4}) occurs at $\theta^*=\tan^{-1} \left(\frac{v^I}{v^R}\right)$  and hence the term $\left(v^R \sin \theta - v^I \cos \theta\right)$ in (\ref{p2cs5}) reduces to zero. Therefore, we have $|v| = |v|_{\mathcal{D} } \sec \delta \le |v|_{\mathcal{D} } \sec \left(\frac{\pi}{P}\right)$.
    	\end{proof}
    \end{lem}
     Hence, the constraints in $P3$ can be written with respect to ${\mathcal{D}}$. The corresponding discrete optimization problem is   
    \begin{align}
    P4:~  &\underset{z, x_i, i \in \Gamma}{\min}~ z \nonumber \\
    ~\text{such that} ~&z \ge \{\F^H\x\}_j^R \cos \theta_p+  \{\F^H\x\}_j^I \sin \theta_p,\nonumber \\
    \theta_p &= (p-1)\frac{2\pi}{P}, ~p = 1, ..., P, ~j=0, ..., N-1 \nonumber \\
    &x_i \in \{-1, +1\} ~\forall~i \in \Gamma \nonumber \\
    &x_i = 0 ~\forall~i \notin \Gamma.
     \label{p2eqn6}
    \end{align}
    Note that $P4$ is only an approximation of $P3$.  Theorem \ref{p2ThmOpt}, stated and proved below, will relate the solution of the approximate problem $P4$ to that of the original optimization problem $P3$.
        \begin{thm}
        	\label{p2ThmOpt}
        	Let $u^*$ be a solution to (\ref{p2eqn5}) and $u^{**}$ be the solution to the corresponding discretized problem (\ref{p2eqn6}). Then, $u^{*} \le  u^{**}\sec\left(\frac{\pi}{P}\right)$.
        	\begin{proof}
        	This is a direct consequence of (\ref{p2cs4}).
        	\end{proof}
        \end{thm}
     $P4$  is a linear integer programming having $(\text{BK}+1)$ optimization variables and $\text{PN}$ linear inequality constraints. This can be solved using techniques such as branch and bound algorithm \cite{beale1965mixed}, which are available with any standard linear integer programming toolbox. $P4$ is computationally more complex than $P1$. However, based on this problem formulation, we propose a heuristic scheme in Section \ref{heur}, which can be implemented in real time.
    \hspace{-1cm}
    \subsection{A Heuristic Algorithm for Reducing PAPR}
    \label{heur}
    A heuristic scheme for solving the PAPR minimization problem $P2$ is given in Algorithm \ref{Alg:PAPR}.  The idea is to iteratively change the sign of each active sub-carriers in succession and check whether PAPR is reduced. Explicitly, in Step 2 of Algorithm \ref{Alg:PAPR}, $+1$ is changed to $-1$, whereas in Step 6, it is done in the reverse direction. If the PAPR is reduced, the change is retained (see function \em{ExchangeSign}\em).  This is done in multiple times until there is no further change in sign is observed.

\begin{center}
	\begin{algorithm}
		\caption{: Heuristic PAPR reduction algorithm}
		\label{Alg:PAPR}
		\begin{flushleft}
			{\bf Inputs:} $\F_{\Gamma}$ - $BK \times N$ partial Fourier Matrix (rows corresponding to position of active indices).  \\
			{\bf Initialization:} $\t =\textbf{1}_{BK \times 1}$ , $\w=\textbf{1}_{BK \times 1}$, $\eta =0.1$. \\
			~~~~~~~~~~~~~~~~~~$v=\max |\F_{\Gamma}^H\t|$, $u= v$. \\
			{\bf Initial Estimation:} $S_{\w} = \sum_{i=1}^{BK} \w_i = BK$.\\
		\end{flushleft}
		\begin{algorithmic}[1]				
			\Repeat 
			\State $(\t, u, v, \w)$ = \em{ExchangeSign}($\F_{\Gamma}, \t, \w, u, v, \eta, 1$)
			 \If {$\sum_{i=1}^{BK} w_i  \ne S_{\w}$}  
			\State   $S_{\w} =  \sum_{i=1}^{BK} \w_i $, $\delta = v$.
			\EndIf
			\State $(\t, u, v, \w)$ = \em{ExchangeSign}($\F_{\Gamma}, \t, \w, u, v, \eta, -1$)			 
			\Until{$\sum_{i=1}^{BK} \w_i ~= S_{\w}$}; 
		\end{algorithmic}
		{\bf Outputs:} $\t$ - Signal vector in the active indices.
	\end{algorithm}
\end{center}
\begin{center}
	\begin{algorithm}
		\caption*{\textbf{Function 1: }ExchangeSign}
		\label{Alg:PAPR_fn}
		\begin{algorithmic}[1]	
			\Function {($\t, u, v, \w$)=\em{ExchangeSign}}	{$\A, \t, \w, u, v, \eta, s$}
			\Repeat   
			\State $\mathbb{I}= \{i: \w_i=\frac{s+1}{2}\}$ 
				 \For {$i \in \mathbb{I}$}
				\State  $\t_{\mathbb{I}_i}= -s$
				\State $v_C = \max |\A^H\t|$
				 \If {$v_C < v$},  
				\State  $v=v_C$, $\w_{\mathbb{I}_i} =\frac{-s+1}{2}$
				  \Else 
				  \State $\t_{\mathbb{I}_i}= s$ 
				 \EndIf
				 \EndFor
			\State $\delta  = u - v$,
			\State $u  = v$.
			\Until{$\delta > \eta v$}	
			\EndFunction		
		\end{algorithmic}
	\end{algorithm}
\end{center}
\par 
It should be noted that the heuristic procedure in Algorithm \ref{Alg:PAPR} may not attain the optimal solution given by solving $P4$. However, it will lead to lower PAPR when compared to the conventional non-coherent OFDM-IM, as demonstrated in the Section \ref{p2sim}. Now, we will derive the ML detector for the proposed schemes.
  \section{Maximum Likelihood (ML) Detection}
  \label{mldet}
For deriving ML detector, we use the same channel model as that of \cite{8091014} given below: 
  \begin{align}
\y = \H\x + \n,
\label{p2eqn9}
  \end{align}
where $\n \sim \CN(0,\sigma^2\I)$ is the additive noise vector and $\H = diag\left(H_0, H_1, ..., H_{N-1}\right)$ is the frequency domain channel matrix with $H_i \sim \CN(0,\sigma_h^2), ~ \forall i=0, ..., N-1$. Theorem \ref{p2ThmDet} will give the optimal detector for the proposed non-coherent OFDM-IM under model (\ref{p2eqn9}).
  \begin{thm}
  	\label{p2ThmDet}
  Under channel model (\ref{p2eqn9}), the ML detector for the proposed PAPR efficient non-coherent OFDM-IM is same as that of the conventional non-coherent OFDM-IM.
  	\begin{proof}
  	Let us consider $j^{th}$ frequency index of the received vector $\y$, i.e., $y_j = H_jx_j + n_j$. In order to prove the theorem, it need only to show that distribution of $y_j$ when $x_j =  \beta e^{j\phi}$ is independent of $\phi$, which is given below.
  	\begin{align}
  	\Pr(y_j|x_j = a) &= \int_{-\infty}^{\infty} \Pr\left(y_j|x_j=a,H_j\right)\Pr\left(H_j\right)dH \nonumber \\
  	&=\CN(0,\sigma^2+|a|^2\sigma_h^2).
  	\label{p2eqn10}
  	\end{align}
  	(\ref{p2eqn10}) is true for $a = \beta e^{j\phi}$ and is independent of $\phi$. This proves the theorem.
  	\end{proof}
  \end{thm}
  \begin{corrol}
  	\label{corrolindxerr}
The probability of index error for the proposed schemes is same as that of conventional non-coherent OFDM-IM.
  \end{corrol}
 Hence, from Theorem \ref{p2ThmDet} and Corollary \ref{corrolindxerr}, it can be concluded that the proposed schemes give the same BER performance as that of the conventional non-coherent OFDM-IM.

  \section{Simulation Results}
  \label{p2sim}
\begin{figure}
 	\centering
 	\includegraphics[width = 0.45\textwidth]{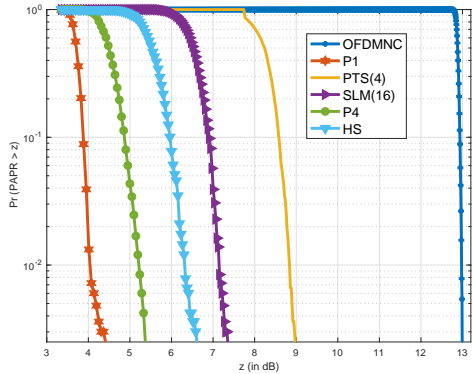}
 	\caption{CCDF of PAPR.}
 	\label{figp1}
 \end{figure}

   A non-coherent OFDM-IM system with $N=128$ sub-carriers, which are split into $B=8$ clusters each having $L=16$ sub-carriers. In each of the clusters. $K= 3$ sub-carriers are made active. In this case, $72$ bits can be transmitted per OFDM frame. Accordingly random binary digits are generated as a group of $72$ and are converted into proper sub-carrier mapping by the combinatoric approach of \cite{6587554}. The PAPR is computed for an oversampling factor $R=4$. We compared the following PAPR reduction techniques.
   \begin{itemize}
   	\item Non-linear optimization problem ($P1$) - The minimax problem is converted into a constrained non-linear minimization problem and solved using $fmincom$ function in $MATLAB2017b$ with default parameters.
   	\item Approximate linear integer program ($P4$) - This is solved using $intlinprog$ function in $MATLAB2017b$  with default parameter settings and $P=5$ discrete levels. 
   	\item Heuristic solution (HS) - Algorithm \ref{Alg:PAPR}
   	\item Selected Mapping (SLM) - The phase factor is generated from angles $\left\{0, \frac{\pi}{2}, \pi, \frac{3\pi}{2}\right\}$ and we used $16$ different signals for representing the same information. The dimension of the phase vector is same as that of the number of active sub-carriers. In other words, SLM scheme used is same as the proposed OSLM, except that the set of phase vector is fixed. 
   	\item Partial Transmit Sequence (PTS)  \cite{581025} -  Here, the OFDM frame is split into four blocks and multiply with phase vector, which is generated from angles $\left\{0, \frac{\pi}{2}, \pi, \frac{3\pi}{2}\right\}$. In this case four different phase vectors are used to represent the same information.
   \end{itemize}
   In each case, the PAPR reduced oversampled time series is generated and complimentary cumulative distribution function (CCDF) is computed. This is also compared with the non-coherent OFDM-IM without any PAPR reduction. However, in this case instead of peak, CCDF of second peak-to-average power ratio is plotted, since its PAPR is a constant (Lemma \ref{lempaprnc}). The CCDF for PAPR of various schemes are shown in Fig. \ref{figp1}. It can be seen that the best performance is obtained for the proposed non-linear optimization $P1$. The performance of the integer program $P4$ is approximately $1~dB$ worse than that of $P1$ at $\text{CCDF} = 0.001$, and the corresponding heuristic scheme performs $1~dB$ worse than the integer program. The performance of $SLM$ is $3~dB$ worse than the optimal solution $P1$, whereas PTS performs $1.5~dB$ worse than $SLM$. This is because, there are $16$ different vectors to represent the same information for $SLM$, whereas it is only $4$ for $PTS$. Finally, the proposed non-linear optimization scheme gives $9~dB$ better performance as compared to conventional non-coherent OFDM-IM. 
 
     	\begin{table}[h]
     		\begin{center}
     				\caption{Comparison of computational time}
     			\begin{tabular}{|l|l|l|l|l|l|}
     				\hline 
     				Scheme & $P1$	& $P4$ &  HS &  SLM ($16$) &  PTS ($4$)  \\
     				\hline 
     				Time (s)  & 22.91 &  719.34 & 0.0025 & 0.0128 & 0.0008\\
     				\hline
     			\end{tabular}		
     			\label{p2table1}
     		\end{center}
     	\end{table}
     	
TABLE \ref{p2table1} shows the average running time (in seconds) of the different schemes in a $i7$ $3.6~GHz$ processor. The integer program $P4$ is taking the most running time, whereas the non-linear program $P1$ gives a better solution than $P4$ with approximately $30$ times faster. However, the heuristic scheme developed from $P4$ provide better PAPR reduction than $SLM$ with $16$ signals and is also computationally efficient. Since, $PTS$ has only four different phase vectors, its running time is the lowest. Therefore, It can be concluded that the proposed heuristic approach is a trade off solution considering performance and complexity.


  \section{Conclusions}
  \label{p2conc}
  We proposed an optimized SLM for PAPR reduction in non-coherent OFDM-IM. We formulated an optimization problem for minimizing PAPR.  The simulation results show that the proposed OSLM scheme has significant PAPR reduction compared to other schemes.  It has also been shown that the heuristic scheme is able to achieve a lower PAPR than SLM and is useful for real time implementation.  Furthermore, the proposed schemes does not have any degradation in BER performance compared to the conventional OFDM-IM scheme. 
  
    
 \bibliography{papr_ofdmim_nc} 
 \bibliographystyle{ieeetr} 

\end{document}